\documentclass[11pt]{article}
\usepackage{graphicx}
\usepackage{amsmath}
\usepackage{amssymb}
\usepackage{amsxtra}
\usepackage[colorlinks]{hyperref}

\title{Cosmological accumulation of conserved numbers in Kaluza-Klein theories}
\author{V.V. Nikulin\\National Research Nuclear University MEPhI\\N-Valer@yandex.ru}

\begin{document}
\maketitle

\begin{abstract}
We develop a new mechanism for the accumulation of conserved numbers in the early Universe in Kaluza-Klein-like theories. The relaxation of the primordial extra space perturbations existing in the early Universe leads to the establishment of a symmetric final state and the appearance of Killing vectors. As a result, the initial non-zero value of symmetry associated numbers occurs after the inflation. We show this conceptual idea on a toy model of 2-dimensional apple-like extra space with U(1) symmetry. This mechanism naturally arises in the Kaluza-Klein theories and can be used to explain the observed cosmological baryon asymmetry.
\end{abstract}

\hspace*{1.8cm} asymmetry, $f(R)$-gravity, cosmological inflation.

\hspace*{1.1cm} 11.30.Fs, 11.30.Ly, 11.30.-j

\section{Introduction}

One of the advantages of using Kaluza{--}Klein compact extra dimensions is that they can explain the origin of internal symmetry in particle physics. The idea of the approach is that the internal symmetry of the gauge theory is considered as a consequence of the geometric properties of compact extra space, characterized by the presence of Killing vector fields \cite{Blagojevic}.

The stability of the compact extra space is the well-known issue of the Kaluza{--}Klein theory. The stabilization can usually be achieved by introducing external material fields \cite{2002PhRvD..66b4036C} or by modifying action for gravity \cite{2006PhRvD..73l4019B}. The process of stabilization obviously should take place in a very early Universe at the energy scales $\sim1/r_0$, when $r_0$ is the radius of compact extra space. In our work \cite{nikulin2019inflationary} we show how dramatically the presence of compact extra space can affect the cosmological inflationary process.

In this paper, we investigate the process of relaxation of the extra space metric during the cosmological inflation. As a result of symmetrization, Killing vector fields appears at the end of inflation and its Noether-associated numbers is asymptotically conserved. The initial non-zero value of this conserved numbers is caused by the extra metric perturbations that took place during inflation. This mechanism could be an explanation for the observed cosmological baryon asymmetry.

\section{Theoretical description}

Today we do not really understand how a compact extra space can be born in higher-dimensional theories. However, we have no reason to believe that its geometry has any symmetry, as this process is clearly random. As a result of further development, the metric of extra space undergoes relaxation and symmetrization. The deep causes for the inevitable appearance of symmetry in this process is related to the establishment of thermodynamic equilibrium and entropy growth \cite{Kirillov:2012gy}.

\subsection{Conserved numbers in Kaluza-Klein theory}

We know that according to the Noether theorem symmetries lead to the conservation of associated numbers. In particular, for (extra) spatial symmetries, the conserved numbers can be interpreted as the physical (angular) moments carried by material fields along the corresponding Killing vectors \cite{Blagojevic,cianfrani2005gauge}.

Spatial symmetry (extra spatial in our case) usually characterized by Killing vector field $\xi^a(x)$. It means that Lie derivative of the extra space metric along the Killing vector field $L_{ \xi}g_{d,mn}=0$ and the metric stays invariant under the small shifts $x_m\rightarrow x_m+\xi^m(x)$. From to the Noether’s theorem (see technical details in \cite{Blagojevic}) we get a conserved current associated with the invariance $\partial_a J^a=0$. This current for any material field $\chi$ is
\begin{equation}
J^a=\frac{\partial L_\text{m}(\chi)}{\partial(\partial_a \chi)}\xi^b\partial_b\chi - \xi^a L_\text{m}(\chi)\,,
\end{equation}
where $L_m$ is a matter Lagrangian.
The associated conserved number
\begin{equation}\label{numb}
Q =\int J^0 \sqrt{|g_4|} \sqrt{|g_d|}\, d^3x\,d^dy\,.
\end{equation}
we can interpret as some component of (angular) momentum.
Until the extra metric reaches a symmetrical final configuration, this number will not be conserved ($Q=Q(t)$). The number will accumulate over time, until the relaxation processes stop. We need to simulate the extra metric and scalar field evolution to the final stable state in order to calculate the value of the accumulated number.

\subsection{Gravitational dynamics of compact space}

Consider as a final result of the stabilization a compact 2{--}dim apple{--}like extra space. This configuration is stationary as was shown in the works \cite{2017JCAP...10..001B,rubin2020cosmology}. It has rotational symmetry which we interpret (in 4{--}dim limit) as $U(1)$ global symmetry with the associated conserved number. In contrast to the one-dimensional circular extra space (which have zero Ricci scalar \cite{sarkar2007particle}) our configuration can lose the symmetry in early high-energetic Universe due to the metric perturbations.

To stabilize the considered extra space, the modified $f(R)${--}gravity is used. First, the higher-dimensional action is taken in the form
\begin{eqnarray}\label{act1}
S = \frac{m_D ^{D-2}}{2}\int d^{D}Z \sqrt{|G|}\left[f(R)+L_\text{m}\right]\, , \quad f(R) = aR^2 + R+c\,.
 \end{eqnarray} 
Here $D=d+4$, $m_D$ is  fundamental $D${--}dimensional Planck mass and $L_m$ is a matter Lagrangian. A conserved number is accumulated in material fields during the stabilization of extra space. We will consider  the simplest case of matter --- massive scalar field:
\begin{eqnarray}
L_\text{m}=\frac{1}{2} G^{MN}\partial_M\chi \partial_N\chi-V(\chi)\,,\quad
V(\chi)=\frac{1}{2}m^2\chi^2\,.
\end{eqnarray}

Consider a $D=d+4${--}dimensional manifold with metric
\begin{equation}\label{metric}
ds^2 = G_{MN}dZ^M dZ^N = g_{\mu\nu}(x)dx^{\mu}dx^{\nu} + g_{d,mn}(x,y)dy^m dy^n \, ,
\end{equation} 
here the metrics $g_{\mu\nu}(x)$ and $g_{d,mn}(x,y)$ corresponds to the $M_4$, $K$ subspaces respectively.
We will consider $M_4$ as a common 4{--}dim space and $K$ as $d${--}dim compact extra space. The signature of D-dim metric is (+ - - - ...) and the Greek indices $\mu, \nu =0,1,2,3$ refer to common 4{--}dim coordinates. Latin indices $m,n = 4, ..., d+3$ refer to the extra coordinates. We will use the following conventions for the Riemann tensor: $R_{ABC}^D=\partial_C\Gamma_{AB}^D-\partial_B\Gamma_{AC}^D+\Gamma_{EC}^D\Gamma_{BA}^E-\Gamma_{EB}^D\Gamma_{AC}^E$ and for the Ricci tensor $R_{MN}=R^A_{MAN}$. We also use unit system $\hbar = c = 1$.

A time evolution of the metric $G_{MN}(x,y)$ is determined by the $f(R)$ Einstein's equations and depends on initial conditions. The dissipation of energy into the 4-dim part of space $M_4$ leads to the decrease of entropy in the compact part of space $K$, as was shown in \cite{Kirillov:2012gy}. Ar a result, a friction term appears, which stabilizes the extra metric $g_{d,mn}(x,y)$. In addition, the inflationary expansion strongly smooths inhomogeneity of 4-dim space:
\begin{equation}\label{uniform}
g_{d,mn}(x,y)
\xrightarrow{t\rightarrow\infty}g_{d,mn}(t,y)\,.
\end{equation}
Time evolution of the extra space was discussed within the Einstein's gravity and Kaluza{--}Klein cosmology framework \cite{Abbott:1984ba}. If a gravitational action has nonlinear Ricci scalar terms -- $f(R)$, the extra metric $g_{d,mn}$ have asymptotically stationary configurations \cite{2006PhRvD..73l4019B,Kirillov:2012gy}:
\begin{equation}\label{statio}
g_{d,mn}(t,y)\rightarrow g_{d,mn}(y).
\end{equation}
See \cite{2002PhRvD..66b4036C,2002PhRvD..66d5029N} for more information.

For simplicity, we can assume that 4-dim space has just de-Sitter metric during inflation
\begin{equation}\label{H}
g_{\mu\nu} = diag(1,-e^{2Ht},-e^{2Ht},-e^{2Ht})\, ,
\end{equation}  
where $H$ is inflationary Hubble parameter. 
The dynamics of inflaton field is not considered here.

To find the stationary configurations of extra space we will use the $f(R)$ Einstein equations:
\begin{equation}\label{eqn}
R_{MN} f' -\frac{1}{2}f(R)g_{d,MN} 
+ \nabla_M\nabla_N f' - g_{d,MN} \square f' = \frac{1}{m_D^{D-2}}T_{MN}.
\end{equation}
Here $\square$ is the d'Alembertian
\begin{equation}\label{dalamb}
\square =\frac{1}{\sqrt{|G|}}\partial_M ( G^{MN}\sqrt{|G|}\partial_N)\,.
\end{equation}
And the contribution of matter is determined by stress{--}energy tensor $T_{MN}$:
\begin{equation}\label{TEI}
T_{MN} = -2 \frac{\partial L_\text{m}}{\partial G^{MN}} + G_{MN} L_\text{m} \, .
\end{equation}

We assume that postulated 4-dim part of metric $g_{\mu\nu}$ \eqref{H} satisfies the higher-dimensional Einstein equations. Next, we will assume that scalar field only depends only on the extra coordinates. It is a result of smoothing out the inhomogeneities of the 3-dim space during inflation. Equation of motion for scalar field $\chi(x,y)=\chi(y)$ is
\begin{equation}\label{mout}
\square_d\chi = - V'(\chi) \, ,
\end{equation}
where $\square_d$ is extra dimensional part of d'Alembertian. 

The very end of the process of forming a compact extra space can be considered as the relaxation of small perturbations of the metric over a stable symmetric vacuum configuration.

\section{Numerical simulation}

\subsection{Vacuum stationary configuration}
As a compact extra space \eqref{uniform}, we take a 2-dimensional sphere-like manifold with the metric
\begin{eqnarray}\label{2dmetr}
g_{2,mn}=\left(\begin{array}{cc}
-r^2e^{2\beta(t,\theta,\phi)}&0\\
0&-r^2e^{2\beta(t,\theta,\phi)}\sin^2{\theta}\\
\end{array}\right)\,.
\end{eqnarray}
where $r$ is characteristic radius of the compact space and the $\beta(t,\theta,\phi)$ is the parameterization function for extra geometry.

To begin with, we will find a vacuum stationary symmetric configuration, which will be the final stage in the evolution of extra space $\beta(t,\theta,\phi)=\beta_\text{st}(\theta)$ and for the scalar field $\chi(t,\theta,\phi)=\chi_\text{st}(\theta)$.
The extra metric has rotational U(1) symmetry associated the presence of Killing vector. The Killing vector field is directed along the polar coordinate $\phi$. The Noether number associated with this U(1) symmetry can be interpreted as the internal polar angular momentum. A similar configuration is used for example in \cite{2017JCAP...10..001B}.


\vspace{1cm}
\begin{figure}[ht!]
\center{\includegraphics[width=\textwidth]{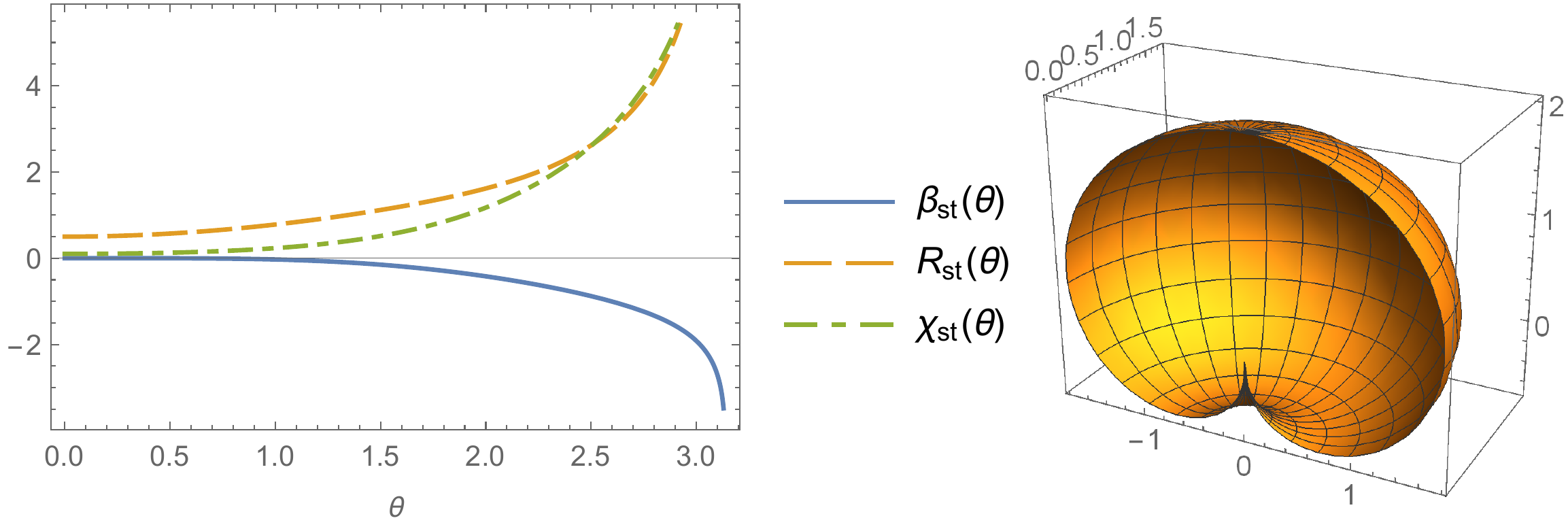}} 
\caption{A typical result of modeling a stationary configuration satisfying the $f(R)$ Einstein equations \eqref{eqn}. On the left: plot of the geometry parameterization function $\beta_{\text{st}}$, the scalar curvature $R_{\text{st}}$ and the material scalar field $\chi_{\text{st}}$ on the azimuthal angle $\theta$ of compact space. On the right: visualisation of the final "apple-shape" stationary configuration of compact 2-dim manifold with metric \eqref{2dmetr}.}
\label{Stationary}
\end{figure}

\newpage
\subsection{Symmetrization process}

Further, to consider the final stage of the relaxation process, we will simulate small perturbations of the metric parameter, scalar curvature, and material scalar field over the stable symmetric state calculated in the last paragraph:
\begin{align}\label{dchi}
\beta(t,\theta,\phi) &=\beta_\text{st}(\theta)\,+\delta\beta(t,\theta,\phi)
\,,\quad \, \delta\beta(t,\theta,\phi) \ll \beta_\text{st}(\theta) \, , \nonumber\\
R(t,\theta,\phi) &=R_\text{st}(\theta)+\delta R(t,\theta,\phi)
\,,\quad \delta R(t,\theta,\phi) \ll R_\text{st}(\theta) , \\
\chi(t,\theta,\phi) &=\chi_\text{st}(\theta)\, +\delta\chi(t,\theta,\phi)
\,,\quad \, \delta\chi(t,\theta,\phi) \ll \chi_\text{st}(\theta) \, . \nonumber
\end{align}

\begin{figure}[ht!]
\center{\includegraphics[width=\textwidth]{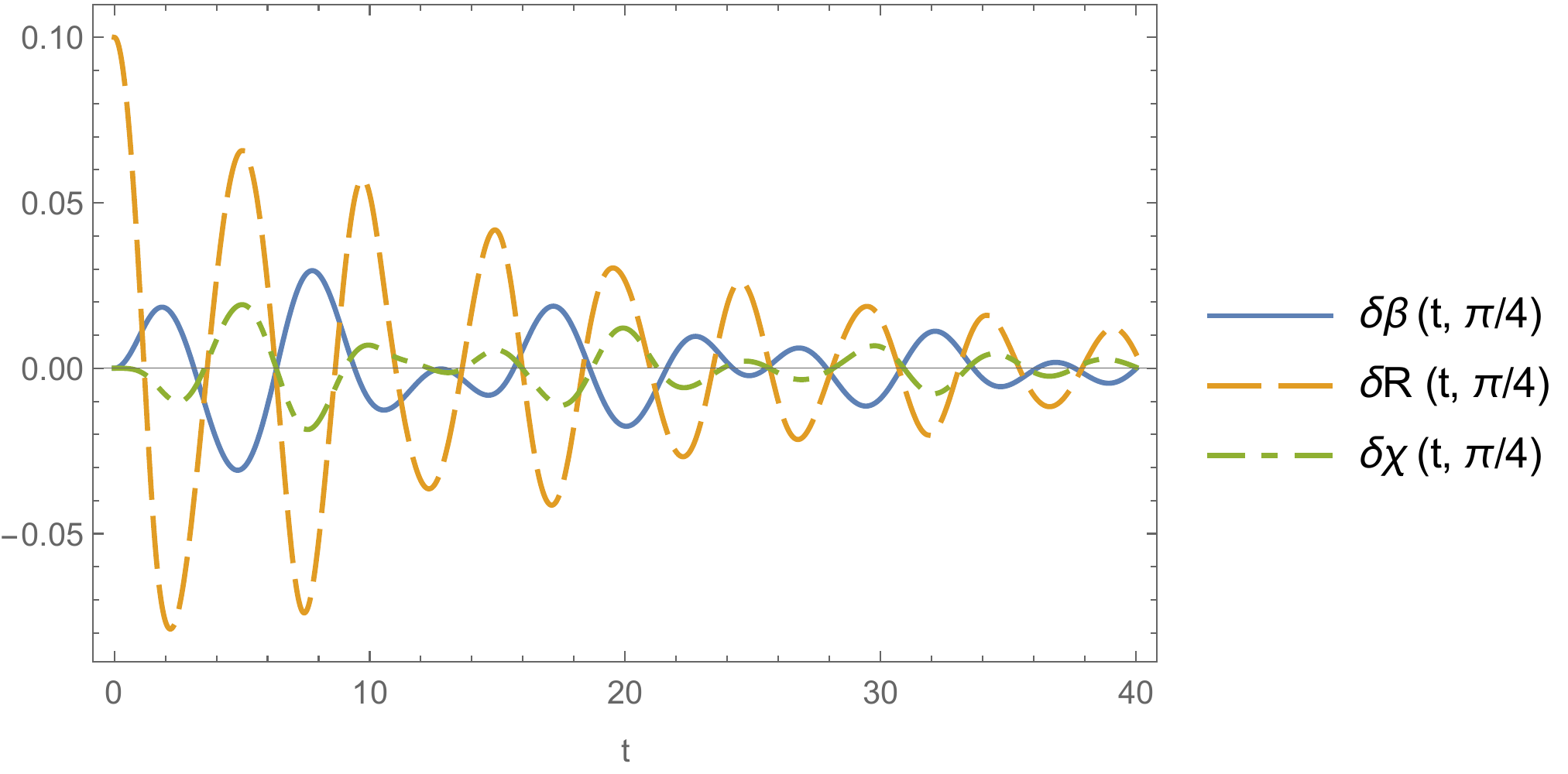}}
\caption{A typical evolution of perturbations $\delta\beta$, $\delta R$, $\delta\chi$ over the stable solution calculated in previous paragraph. As an example, the behavior of the polar mode $n=2$ is shown (standing wave along $\phi$ coordinate). Oscillations are taken at a point $\theta=\pi/4$, at other points damping behaves similarly.}
\label{Osc}
\end{figure}

By linearizing the Einstein's equations \eqref{eqn}, and solving it \cite{nikulin2020formation} for natural random initial conditions, we obtain damped oscillations, which are shown in Fig.\ref{Osc}. The dumping occurs for all angles $\theta$ which shows the stability of the resulting configuration. This is due to the friction term commonly generated in the de Sitter space. The latter leads to the final stabilization to the U(1) symmetric extra space configuration.

\vspace{1cm}

\subsection{Initial accumulation of U(1) number}

After the end of the relaxation processes shown in the previous subsection, a symmetric U(1) configuration is achieved. The U(1)-number associated with the Noether theorem \eqref{numb} will now be conserved. But in this section we are interested in how this number $Q$ could have accumulated initially, until the end of the relaxation and symmetrization processes. The perturbed solutions simulated earlier allow us to compute $Q(t)$ number. In the accompanying volume we get (from \eqref{numb},\eqref{metric},\eqref{dchi}):
\begin{align}\label{numb2}
Q(t)&=\int \partial^0 \chi \partial_\phi \chi\, r^2 e^{2\beta} \sin \theta\, d\theta d\phi = \\
&= \int \partial^0 \delta\chi(t,\theta,\phi) \partial_\phi \delta\chi(t,\theta,\phi)\, r^2 e^{2\bigl(\beta_{st}(\theta)+ \delta\beta(t,\theta,\phi)\bigr)} \sin \theta\, d\theta d\phi \, . \nonumber
\end{align}

\begin{figure}[ht!]
\center{\includegraphics[width=0.9\textwidth]{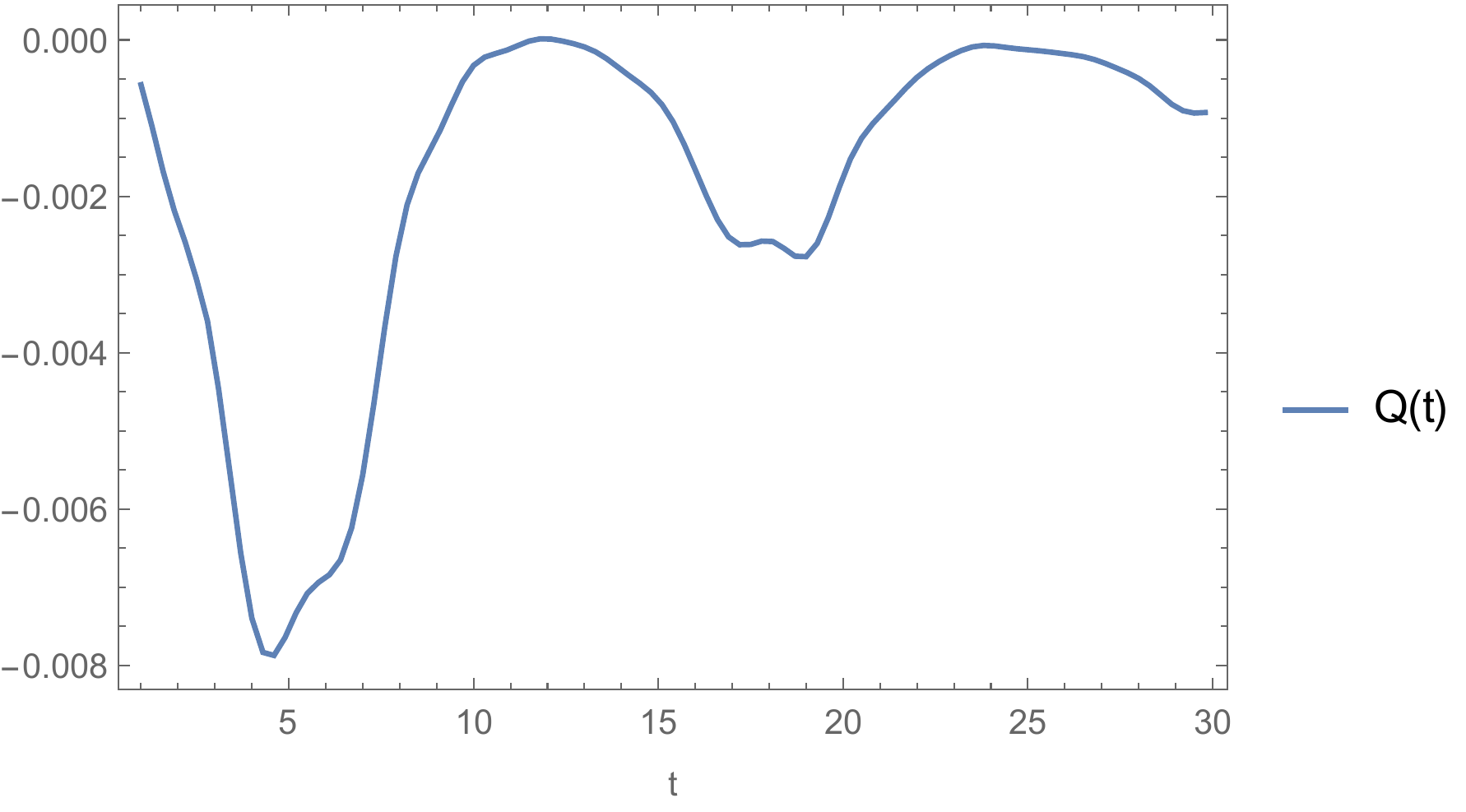}} 
\caption{Typical time evolution of the U(1) number $Q(t)$ during the symmetrization of compact extra space. The number calculated numerically from \eqref{numb2}.}
\label{Charge}
\end{figure}

\vspace{0.5cm}
The end of inflationary process have very rapid transition to the reheating stage via the violation of the slow-roll conditions. Due to this the extra metric is quickly symmetrized (for $H\lesssim 1/r$ extra space perturbations are rapidly suppressed), while the scalar field go into the oscillating mode. After the inflation, stationary extra metric $\beta(t,\theta,\phi) = \beta_{st}(\theta)$ give us the equation of motion for matter \eqref{mout} with nonperturbed symmetrical d'Alambertian. As a result, Noether's theorem starts to be fulfilled and $Q$ ceases to depend on time. Traveling waves of the scalar field, carrying an internal angular momentum is now permanently enclosed inside extra space, since the number $Q$ is now conserved. The initially accumulated $Q(t)$ will now remain unchanged. The Universe enters the hot stage with a nonzero initial value of U(1) global conserved number.

\section{Conclusion}

In this research we show how the dynamics of compact extra space leads to a nonzero initial accumulation of some conserved number. Such gravitational dynamics of compact extra metric should naturally occur in the early ($H \gg 1/r$) higher-dimensional Universe. The stabilization of the extra metric lead to a symmetrical stationary final configuration. We considered the case of a final U(1) rotationally symmetric state with corresponding conserved number.

Such an accumulation mechanism arising in Kaluza-Klein theories can be used to explain the origin of the cosmological baryon asymmetry \cite{2006PZETF...83..3,2009NuPhB.807..229D}. It is known that the baryon number is described by the global $U(1)${--}symmetry. In Kaluza-Klein theories it could be realized as the rotational symmetry of the 2{--}dim compact extra space \eqref{2dmetr}. However, to transfer the baryon number, additional interaction term between the fermion and the scalar field is required (for details, see work \cite{1993PhRvD..47.4244D}).

In future works, we plan to develop a Kaluza-Klein mechanism for transferring asymmetry into the fermions in order to explain specifically the cosmological baryon/lepton asymmetry.

\section*{Acknowledgements}
The work was supported by the Ministry of Science and Higher Education of the Russian Federation, Project "Fundamental properties of elementary particles and cosmology" No 0723-2020-0041.




\bibliographystyle{abbrv}
\bibliography{bibl}

\begin{thebibliography}{10}

\bibitem{Abbott:1984ba}
R.~B. Abbott, S.~M. Barr, and S.~D. Ellis.
\newblock {Kaluza-Klein Cosmologies and Inflation}.
\newblock {\em Phys. Rev. D}, D30:720, 1984.

\bibitem{Blagojevic}
M.~Blagojevi{\'c}, F.~W. Hehl, and eds.
\newblock {\em {Gravitation and Gauge Symmetries}}.
\newblock Institute of Physics Publishing, Bristol, 2002.

\bibitem{2017JCAP...10..001B}
K.~A. {Bronnikov}, R.~I. {Budaev}, A.~V. {Grobov}, A.~E. {Dmitriev}, and S.~G.
  {Rubin}.
\newblock {Inhomogeneous compact extra dimensions}.
\newblock {\em Journal of Cosmology and Astroparticle Physics}, 10:001, 2017.

\bibitem{2006PhRvD..73l4019B}
K.~A. {Bronnikov} and S.~G. {Rubin}.
\newblock {Self-stabilization of extra dimensions}.
\newblock {\em Phys. Rev.~D}, 73(12):124019, 2006.

\bibitem{2002PhRvD..66b4036C}
S.~M. {Carroll}, J.~{Geddes}, M.~B. {Hoffman}, and R.~M. {Wald}.
\newblock {Classical stabilization of homogeneous extra dimensions}.
\newblock {\em Phys. Rev.~D}, 66(2):024036, 2002.

\bibitem{cianfrani2005gauge}
F.~Cianfrani, A.~Marrocco, and G.~Montani.
\newblock Gauge theories as a geometrical issue of a kaluza--klein framework.
\newblock {\em International Journal of Modern Physics D}, 14(07):1195--1231,
  2005.

\bibitem{1993PhRvD..47.4244D}
A.~{Dolgov} and J.~{Silk}.
\newblock {Baryon isocurvature fluctuations at small scales and baryonic dark
  matter}.
\newblock {\em Phys. Rev.~D}, 47:4244--4255, 1993.

\bibitem{2009NuPhB.807..229D}
A.~D. {Dolgov}, M.~{Kawasaki}, and N.~{Kevlishvili}.
\newblock {Inhomogeneous baryogenesis, cosmic antimatter, and dark matter}.
\newblock {\em Nucl. Phys.~\bf{B}}, 807:229--250, 2009.

\bibitem{2006PZETF...83..3}
M.~Y. {Khlopov}.
\newblock {Composite Dark Matter from 4-th Generation}.
\newblock {\em Pis'ma Zh. Ehksp. Teor. Fiz.}, 83:3--6, 2006.

\bibitem{Kirillov:2012gy}
A.~A. Kirillov, A.~A. Korotkevich, and S.~G. Rubin.
\newblock {Emergence of symmetries}.
\newblock {\em Phys. Lett.}, B718:237--240, 2012.

\bibitem{2002PhRvD..66d5029N}
S.~{Nasri}, P.~J. {Silva}, G.~D. {Starkman}, and M.~{Trodden}.
\newblock {Radion stabilization in compact hyperbolic extra dimensions}.
\newblock {\em Phys. Rev.~D}, 66(4):045029, 2002.

\bibitem{nikulin2020formation}
V.~V. Nikulin, P.~M. Petriakova, and S.~G. Rubin.
\newblock Formation of conserved charge at the de sitter space.
\newblock {\em Particles}, 3(2):355--363, 2020.

\bibitem{nikulin2019inflationary}
V.~V. Nikulin and S.~G. Rubin.
\newblock Inflationary limits on the size of compact extra space.
\newblock {\em International Journal of Modern Physics D}, 28(13):1941004,
  2019.

\bibitem{rubin2020cosmology}
S.~G. Rubin.
\newblock Cosmology and matter-induced branes.
\newblock {\em Symmetry}, 12(1):45, 2020.

\bibitem{sarkar2007particle}
U.~Sarkar.
\newblock {\em Particle and Astroparticle physics}.
\newblock CRC Press, 2007.

\end{thebibliography}

\end{document}